\documentclass[aps,prd,twocolumn,superscriptaddress,groupedaddress,nofootinbib]{revtex4-2}
\usepackage[colorlinks=true,citecolor=blue,linkcolor=red,urlcolor=red]{hyperref}
\usepackage[sort&compress]{natbib}
\usepackage{graphicx}
\usepackage{amssymb}
\usepackage{braket}
\usepackage{amsmath}
\usepackage{hyperref}
\usepackage{dsfont}
\usepackage{bm} 
\usepackage{physics}
\usepackage{tabularx}
\usepackage{verbatim}
\usepackage{float}
\usepackage{romannum}
\usepackage{bbold}
\usepackage{mathtools}

\renewcommand\k{{\bf k}}
\renewcommand\r{{\bf r}}

\newcommand\x{{\bf x}}

\begin{document}
	
\author{Anirudh Gundhi}
\email{anirudh.gundhi@units.it}
\affiliation{Department of Physics, University of Trieste, Strada Costiera 11, 34151 Trieste, Italy}
\affiliation{Istituto
	Nazionale di Fisica Nucleare, Trieste Section, Via Valerio 2, 34127 Trieste,
	Italy}

\title{Decoherence due to the Casimir effect?}
	
\date{\today}

\begin{abstract}
 Open system dynamics of an electron is studied in the presence of radiation field, confined between two parallel conducting pates. It has  been suggested in previous works that the quantized zero-point modes of this field lead to finite decoherence effects, possibly due to the Casimir force. However, in this work it is shown that the decoherence found in previous works is due to the sudden switching on of the system-environment interaction and due to the acceleration of the electron enforced by the background paths whose superposition was analyzed.
  The work discusses important theoretical aspects of the setup and shows that while coherence might be lost due to bremsstrahlung induced by an external or the image potential, it cannot be lost due to the mere presence of the quantum vacuum fluctuations between the plates.
\end{abstract}

\maketitle

\section{Introduction}
Previous works \cite{Caldeira1991,Santos1994, Diosi1995, Ford1997, BandP, Unruh_Coherence, Gundhi:2023vjs} have considered the intriguing possibility of observing decoherence effects  solely due to the presence of vacuum fluctuations.\footnote{An introduction to open quantum systems can be found in \cite{KieferDecoherence, Schlosshauer2007} See also \cite{Kiefer1992} for a discussion on decoherence in electrodynamics.} Any such loss of coherence would also compete with the models of wavefunction collapse \cite{Bassi2003,Bassi2013,Arndt2014,Carlesso2022} in predicting the quantum-to-classical transition of perfectly isolated systems. However, it was shown in \cite{Diosi1995, Unruh_Coherence, Gundhi:2023vjs} that there can be no observable decoherence due to vacuum fluctuations alone. The overall effect of the vacuum is merely to dress the bare electron with a cloud of virtual photons \cite{Diosi1995,Unruh_Coherence}. The QED vacuum state depends upon the position of the charged particle but has no \textit{memory} of its trajectory. It cannot acquire which-path information about the particle and results in no irreversible loss of coherence.  Therefore, the presence of zero-point modes  cannot be deduced by measuring the loss of coherence for the free electron.

One can infer the presence of vacuum fluctuations, however, by measuring the stress two parallel and perfectly conducting plates exert on each other \cite{Casimir1948, Sabisky1973, Lamoreaux1997, LamoreauxE1998, Lamoreaux1999}. The Simplest way to understand the Casimir effect is to notice that the boundary conditions imposed by the plates do not allow for all the zero-point modes to be present, since the allowed mode wavelengths are constrained by the separation between the plates. Thus, there is a favorable energy gradient in bringing the plates from infinity to a finite separation, resulting in the attractive Casimir force. A review of the subject can be found in \cite{Spruch1993,Milton2001}.

The question then arises whether the Casimir force  can also \textit{monitor} the motion of a charged particle and thus lead to decoherence. In other words, it remains to be understood whether the role of the environment played by vacuum fluctuations changes from being trivial, in empty space, to something significant and observable, when the radiation field is confined between conducting plates. Such effects have been studied in  \cite{Ford1993,Ford2004,Mazzitelli2003} (single conducting plate) and in \cite{Hsiang2006} (where the case of parallel conducting plates was also considered explicitly). It is concluded in \cite{Ford1993,Mazzitelli2003,Hsiang2006} that the zero-point fluctuations of the radiation field result in some decoherence at the level of the electron, and that this effect might be related to the Casimir force. 

Instead, in this work it will be shown that decoherence found in previous works is due to the sudden switching on of the interaction between the electron and the radiation field \cite{Ford1993}, and due to the acceleration of the electron along the trajectories whose superposition is analyzed \cite{Mazzitelli2003,Hsiang2006}. This loss of coherence should not be interpreted due to the mere presence of the vacuum fluctuations.  There is no physical mechanism through which the electron's which-path information can be lost irreversibly to the environment, especially if the plates are treated as ideal conductors and effects such as ohmic resistance to the image currents, as considered in \cite{Anglin1996}, are ignored.\footnote{Non-ideal conductors \cite{Levinson2004} or dynamical boundary conditions \cite{Dalvit2000,Wu2005} might also lead to irreversible decoherence. See also \cite{Sinha2020} for open system dynamics of a polarizable dielectric particle.} 
Nevertheless, the initial \textit{jolt}, resulting from the sudden switching on of the interaction, leads to peculiar features in the reduced density matrix which are typically not found in other open systems. These features are explained by realizing that part of the initial jolt in the radiation field would  affect the dynamics even at later times, since it is reflected back and forth by the conducting plates on either side of the electron. Although it would have little observational relevance in typical scenarios, the dynamics might still serve as a special reference for this widely discussed technical aspect of open quantum systems, dealing with the transient effects due to the sudden switching on of the system-environment (S-E) interaction \cite{Hakim1985,SKODJE1988,Grabert1988,UnruhZurek1989,HuPazZang1992,Costa2000,Das2015,Agon2017}.

 Nevertheless, from a physical point of view in which such a jolt would be absent,  decoherence can only result from bremsstrahlung induced by the Coulomb potential of the image charges or an external potential, but not from the quantum vacuum fluctuations alone.

\section{Image charges}\label{Sec:ImageCharges}
\begin{figure}[!ht]
\centering
\includegraphics[width=0.70\linewidth]{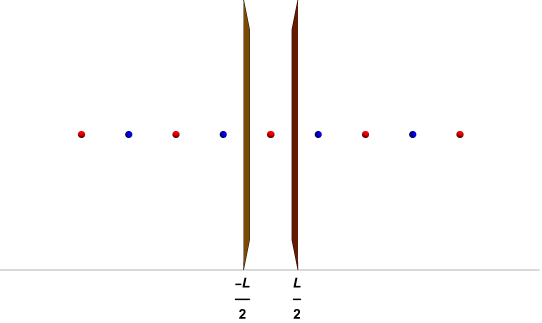}
\caption{Infinite number of positive (blue) and negative (red) image charges induced by the plates, located at $x=-L/2$ and $x=L/2$, to cancel the Coulomb field of the electron (center) along their surface.}\label{Images}
\end{figure}
The electric field along the surface of an idealized perfect conductor is zero. It is well known from standard electrostatics that if an electron is placed between two parallel conducting plates, as shown in Fig.~\ref{Images}, this requirement can be met with the help of  an infinite number of image charges, whose positions are determined by the position of the electron between the plates. If the electron is placed at a coordinate $x$, the position state of the image charges is given by
\begin{align}\label{AEQ:ImageLocation}
\ket{{\color{blue}{+}}}_x &=\ket{ {\color{blue}\cdots},-(2n-1)L-x,{\color{blue}\cdots},(2n-1)L- x,{\color{blue}\cdots}}\,,\nonumber\\
\ket{{\color{red}{-}}}_x &= \ket{{\color{red}\cdots},-2nL+x,{\color{red}\cdots},2nL+ x,{\color{red}\cdots}}\,,
\end{align}
where $n\geq 1$ is an integer. 

From an open quantum system point of view, these infinite image charges belong to the environment. One might therefore ask whether or not there would be any direct positional decoherence due to the image charges themselves, since the position state of the images is perfectly correlated to the position state of the electron.  This is the subject of this section, while decoherence due to vacuum fluctuations is studied in the following ones. 

To find an answer for the loss in coherence, one might look at the off-diagonal elements of the reduced density matrix $\hat{\rho}_r$. The standard expression for $\hat{\rho}_r$ is given by 
\begin{align}\label{AEQ:RedDensityDef}
\hat{\rho}_r(t) = \Sigma_{E}\bra{E}\ket{\Psi_t}\bra{\Psi_t}\ket{E}\,,
\end{align}
where $\ket{\Psi_t}$ represents the full S-E state at time $t$, and $\ket{E}$ the environmental basis states. In general, due to the S-E interaction, the state of the environment $\ket{\mathcal{E}}$ becomes correlated to the position of the system such that $\ket{\Psi_t} = \int dx\psi_t(x)\ket{\mathcal{E}_x}\ket{x}$, where $\psi_t(x)$ is the probability amplitude for the system to be at a position $x$ at time $t$. In such a scenario,  the off-diagonal elements of $\hat{\rho}_r$ are given by
\begin{align}
\bra{x}\hat\rho_r(t)\ket{x'} = \bra{\mathcal{E}_{x'}}\ket{\mathcal{E}_x}\psi_t(x)\psi^{*}_t(x')\,.
\end{align}
If the environmental states for two different states of the system are completely orthogonal, the system is generally understood to have decohered completely. If one follows this line of reasoning strictly, the decoherence kernel $\mathcal{D}(x,x')$ due to the image charges will be given by
\begin{align}\label{AEQ:DecKernel}
\mathcal{D}(x,x'):=\bra{\mathcal{E}_{x'}}\ket{\mathcal{E}_x} =\left._{x'}\bra{{\color{blue}{+}},{\color{red}{-}}}\ket{{\color{blue}{+}},{\color{red}{-}}}_x\right.\,.
\end{align}
It is clear that $\mathcal{D}(x,x')=0$ whenever $x'\neq x$ and $\mathcal{D}(x,x')=1$ for $x'=x$. Thus, one might conclude that the image charges measure the electron's position perfectly and instantaneously. 

However, this suppression of the off-diagonal elements represents \textit{false decoherence}. In order to observe the loss of fringe contrast in a double-slit experiment, the environment must be able to acquire which-path information rather than which-position information about the system. The interference fringes appear because the electron reaches a given point on the detector screen while being in a superposition of different paths. As it approaches the screen and is detected at a given point, the state of the image charges would only be correlated to the position at which the electron is detected on the screen, and would not have any \textit{memory} of which slit the electron passed through.\footnote{It might be argued that by \textit{looking} at the images one can deduce which slit the electron passed through, and therefore the interference pattern should be lost. However, in this line of reasoning, an extra  environment is introduced which can look at the images themselves and then remember the outcome. This should not be done, since the only environment we have is that of the image charges.} In other words,  $\mathcal{D}\to 1$ for the two paths that pass through different slits but end at a given point on the screen where the electron is detected. Therefore, there would be no loss in fringe contrast due to the images. The situation here is analogous to empty space where it is understood why there are no decoherence effects due to the particle's Coulomb field itself \cite{Giulini1995,Kiefer2003}. For a loss in fringe contrast, the environment should continuously monitor the system, like a thermal bath, where the photons scattering off the electron at different times  can effectively measure its full trajectory. Instead, if $\mathcal{D}$ depends only on the latest position of the system, which-path information cannot be acquired.

While the image charges cannot directly decohere the electron, there is an indirect way by which they can. It is due to the acceleration caused by the image charges, resulting in bremsstrahlung, which carries away which-path information about the electron. The effective potential at the electron's location can be  computed by summing over the Coulomb potential due to all the infinite images. In doing so, one might first notice that the distance between a given negative image charge and the electron is $|\pm 2nL +x-x| = 2nL$. Since it is independent of the electron's position, the Coulomb potential due to the negative images  cannot influence the electron's dynamics, as it simply adds a constant to the Hamiltonian. It is therefore sufficient to compute  the Coulomb potential due to the positive image charges only. It is given by
\begin{align}
V_{\mathrm{im}} = -\alpha\hbar c\sum_{n=1}^{\infty} \left[\frac{1}{(2n-1)L-2x}+\frac{1}{(2n-1)L+2x}\right]\,,
\end{align}
where $\alpha=e^2/(4\pi\epsilon_{0}\hbar c)$ is the fine structure constant. This sum is divergent, but not the dependence on $x$. To see this, the expression above is rewritten as
\begin{align}
V_{\mathrm{im}} &= -\alpha\hbar c\sum_{n=0}^{\infty} \frac{2(2n+1)L}{(2n+1)^2L^2-4x^2}\nonumber\\
&=-\alpha\hbar c\sum_{n=0}^{\infty}\left[\frac{2}{(2n+1)L}\left(1-\frac{4x^2}{(2n+1)^2L^2}\right)^{-1}\right]\,.
\end{align}
From the setup shown in Fig.~\ref{Images}, it can be seen that ${-L/2<x<L/2}$ and therefore the term $\left(1-4x^2/((2n+1)L)^2\right)^{-1}$ can be written as a geometric series. The zero$^{\mathrm{th}}$ order term in $x$ of this expansion would again just give an irrelevant (divergent) constant contribution to the Hamiltonian and can therefore be ignored. The remaining $x$-dependent part of the potential is given by
\begin{align}\label{AEQ:EffPotential}
&V_{\mathrm{im}}(x) = \frac{-2\alpha\hbar c}{L}\sum_{n=0}^{\infty}\frac{1}{2n+1}\times\frac{4x^2}{(2n+1)^2 L^2 - 4x^2}=\nonumber\\
&\frac{\alpha\hbar c}{2L}\left[H\left(-1/2-x/L\right)+H\left(-1/2+x/L\right)+\ln(16)\right]\,,
\end{align}
where, $H(x)$ is the standard harmonic number function generalized to the domain of real numbers. Note that the expression for the effective potential $V_{\mathrm{im}}$ has been previously derived in \cite{Barton1970}, using a more formal technique, but not by explicitly summing over the Coulomb potential due to all the image charges. 
\begin{figure}[!ht]
\centering
\includegraphics[width=0.70\linewidth]{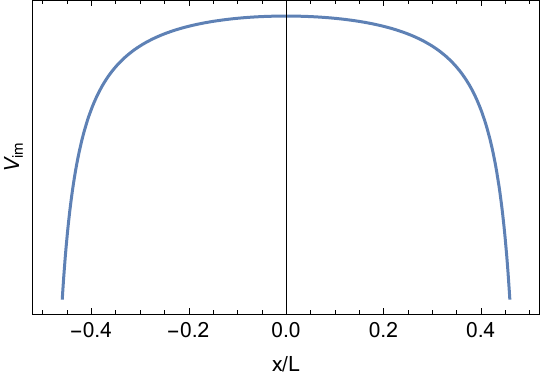}
\caption{Exact Coulomb potential due to the infinite image charges, Eq.~\eqref{AEQ:EffPotential}, at a location $x$ between the plates.}\label{ImagePotential}
\end{figure}

It can be seen from Fig.~\ref{ImagePotential} that the electron would be in an unstable equilibrium at $x=0$ and inevitably attracted to either one of the two plates. This is inevitable, because in quantum mechanics one cannot have both the position and velocity  to be zero simultaneously.\footnote{In this context, one may find the discussion on the quantum mechanical treatment of the inverted oscillator in \cite{Barton1986} relevant.}  The effect of the images on the electron dynamics if far from being negligible. It is clear that the electron would experience strong acceleration, especially near $x=\pm L/2$, and thus decohere due to which-path information being carried away by bremsstrahlung.  To conclude this section, a simple scenario is analyzed in which the effects of the image charges can be computed easily, and contrasted with the predictions in empty space. 

To study decoherence effects over large timescales, one might introduce an external harmonic potential potential,
\begin{align}
V_{\mathrm{ext}}(x) = \frac{1}{2}m\Omega^2x^2\,,
\end{align}
which restricts the electron dynamics in the neighborhood of the origin. This is because   the electron would otherwise run away towards the plates. In such a scenario, it is useful to expand $V_{\text{im}}(x)$ in powers of $x/L$ and to keep the leading order term. The potential due to the images~\eqref{AEQ:EffPotential} can be written as
\begin{align}
V_{\mathrm{im}}(x) = \frac{\alpha\hbar c}{L}\sum_{n=1}^{\infty}\frac{\psi(2n,1/2)}{(2n)!}\left(\frac{x}{L}\right)^{2n}\,,
\end{align}
where $\psi(2n,1/2):=d^{2n}\psi(x)/dx^{2n}$, evaluated at $x=1/2$, $\psi(x)$ being the digamma function. Therefore, for a large enough $\Omega$, the overall effect of the image charges is to lower the frequency of the harmonic oscillator (since $\psi(2,1/2) = -16.8288$)
\begin{align}
V_{\mathrm{eff}} &= V_{\mathrm{im}}+V_{\mathrm{ext}} = \frac{1}{2}m\Omega^2_{\mathrm{eff}}x^2\,,\qquad \nonumber\\
\Omega^2_{\mathrm{eff}}&=\Omega^2+\frac{\alpha\hbar c\psi(2,1/2)}{mL^3}\,,
\end{align}
and thus redshift the spectrum. To leading order in $1/L$, the emission of bremsstrahlung  between the plates is the same as in empty space. This is simply because in the limit $L\to \infty$ we should recover the dynamics as in empty space. Therefore, to leading order, decoherence in the presence of parallel plates can be described by the master equation for bremsstrahlung \cite{Gundhi2024}
\begin{align}\label{AEQ:Brem}
\partial_t\hat{\rho}_r(t) =&-\frac{i}{\hbar}\left[\frac{\hat{p}^2}{2m}+\frac{1}{2}m\Omega^2_{\mathrm{eff}}\hat{x}^2,\hat{\rho}_{r}(t)\right]\nonumber\\
&-\frac{i\alpha\Omega^2_{\mathrm{eff}}}{3mc^2}\left[\hat{x},\{\hat{p},\hat{\rho}_r(t)\}\right] -\frac{\alpha\Omega^3_{\mathrm{eff}}}{3c^2}\left[\hat{x},\left[\hat{x},\hat{\rho}_{r}(t)\right]\right]\,.
\end{align}   
 We see that since $\Omega_{\mathrm{eff}}<\Omega$, the overall effect of the image charges is to increase the coherence length, simply because the plates effectively lower the frequency of the external harmonic potential $V_{\mathrm{ext}}$. Nevertheless, it is clear that the image potential exerts an additional force on the electron, which is absent in empty space, leading to decoherence rates which are different compared to those in empty space. The rate can be lower or higher, depending upon the setup. For other experimental scenarios, one might use the exact potential derived in Eq.~\eqref{AEQ:EffPotential}  and compute the decoherence rate using the full noise kernel, presented in the next section. Further details concerning the effect of image charges on the electron dynamics will not be discussed. The main objective of this section is to highlight how the effective potential inevitably accelerates the  electron, leading to irreversible decoherence. 

\section{Zero-point modes}
The final aspect of the analysis is to understand whether vacuum fluctuations, confined between conducting plates, can lead to observable decoherence. To answer this question, one must study if the suppression of the off-diagonal elements of the reduced density matrix implies genuine decoherence, and therefore check if the environment \textit{remembers} the state of the system, even when its interaction with the system is switched-off after some finite time. If it does not, the general implication would be that the environment is only correlated to the state of the system at a given time, and not to its history/trajectory. The situation would then be identical to that of Sec.~\ref{Sec:ImageCharges}, where it was explained why environments of this type do not lead to a loss in the fringe contrast. 

\subsection{Master equation}
Up to second order in the charge $e$, using the influence-functional formalism \cite{Feynman_Vernon,calzetta_hu_2008}, the time evolution of the reduced density matrix of the electron  was recently derived  in \cite{Gundhi:2023vjs}, starting from the nonrelativistic QED Lagrangian. It reads
\begin{align}\label{AEQ:MasterEquation}
\partial_t\hat{\rho}_r=&-\frac{i}{\hbar}\left[\hat{\mathrm{H}}_s,\hat{\rho}_{r}(t)\right]\nonumber\\  &-\frac{1}{\hbar}\int_{0}^{t-t_i} d\tau \mathcal{N}(t;t-\tau)\left[\hat{x},\left[\hat{x}_{\text{\tiny{H}}_s}(-\tau),\hat{\rho}_{r}(t)\right]\right]\nonumber\\&+\frac{i}{2\hbar}\int_{0}^{t-t_i}
d\tau{D}(t;t-\tau)\left[\hat{x},\{\hat{x}_{\text{\tiny{H}}_s}(-\tau),\hat{\rho}_{r}(t)\}\right]\,.
\end{align}
 In the equation above, ${\mathrm{H}}_{s}$ is the system part of the full Hamiltonian $\text{H}$
 \begin{align}
 \text{H}:= \text{H}_{s}+\text{H}_{\text{\tiny{EM}}}+\text{H}_{\text{int}}\,,
 \end{align}
 where $\text{H}_{\text{\tiny{EM}}}$ governs the free evolution of the radiation field and $\text{H}_{\text{int}}$ governs the S-E interaction, 
 \begin{align}\label{AEQ:Hint}
\text{H}_{\text{\tiny{EM}}}= \frac{\epsilon_0}{2}\int d^3r(\mathbf{\Pi}^2+c^2\mathbf{B}^2)\,,\qquad \text{H}_{\text{int}}=e\r\mathbf{\Pi}(\r,t)\,,
 \end{align}
where $\mathbf{\Pi}:=-\textbf{P}/\epsilon_0$, $\textbf{P}$ being the conjugate momentum of the radiation field. The system Hamiltonian is given by 

\begin{align}\label{eq:Hs}
\hat{\text{H}}_{s}=\frac{\hat{p}^2}{2m}+{V}_{\mathrm{eff}}(\hat{x})+{V}_{\text{\tiny{EM}}}(\hat{x})\,,
\end{align}
where $V_{\mathrm{eff}}$ is the effective potential due to the images and some external potential that might be applied, while $V_{\text{\tiny{EM}}} = e^2\omega^3_{\text{\tiny{max}}}\hat{x}^2/({3\pi^2\epsilon_0c^3})$ is a divergent contribution, which scales with the UV cutoff $\omega_{\text{\tiny{max}}}$, originating from the Legendre transformation relating the QED Lagrangian and the Hamiltonian  \cite{Caldeira1991,Gundhi:2023vjs}. The role played by $V_{\text{\tiny{EM}}}$ is to cancel the divergence coming from the dissipation kernel $D$ in the master equation \eqref{AEQ:MasterEquation}, as shown in \cite{Gundhi:2023vjs}. 

For the motion of the electron considered along the $\x$ axis, the noise kernel $\mathcal{N}$ and the dissipation kernel $D$ in Eq.~\eqref{AEQ:MasterEquation} are defined to be
\begin{align}\label{AEQ:Kernels}
\mathcal{N}(t_1;t_2):=&\frac{e^2}{2\hbar}\left\langle\{ \hat{\Pi}^x(t_1), \hat{\Pi}^x(t_2)\}\right\rangle_0\,,\nonumber\\
D(t_1;t_2):=&\frac{ie^2}{\hbar}\left\langle\left[ \hat{\Pi}^x(t_1), \hat{\Pi}^x(t_2)\right]\right\rangle_0\theta(t_1-t_2)\,.
\end{align} 
In Eq.~\eqref{AEQ:Kernels}, the operator $\hat{\Pi}^x(t)$ is the freely evolved Heisenberg operator, the same as the transverse electric field operator of the free EM field,  while the expectation value $\langle\rangle_0$ is  taken with respect to the initial state of the environment. To study decoherence due to vacuum fluctuations, the initial state of the environment  is taken to be the ground state $\ket{0}$ of the free electromagnetic (EM) field, assuming that there are no photons at the initial time $t_i=0$ and that the initial S-E state is completely uncorrelated, i.e., $\hat{\rho}(t_i) = \hat{\rho}_{s}(t_i)\otimes\ket{0}\bra{0}$. The kernel $\mathcal{N}$ describes decoherence while $D$ describes the energy lost by the system to the environment. Since it is important to  distinguish between the decoherence effects due to bremsstrahlung and those due to zero-point modes of the EM  field, the off-diagonal elements are computed with the electron kept steady in a superposition of $x$ and $x'$, so that decoherence due to bremsstrahlung does not show up. In such a scenario, $\hat{x}_{\text{\tiny{H}}_s}(-\tau)$, which is the Heisenberg time-evolved position operator with respect to the system Hamiltonian $\hat{\text{H}}_{s}$, is given by $\hat{x}-\hat{p}\tau/m$. Moreover, in the absence of acceleration, the dissipation kernel would not give any contribution to the master equation as the electron can only lose energy to the environment via radiation emission upon acceleration (to second order in $e$). 

Thus, the master equation describing decoherence reduces to
\begin{align}\label{AEQ:NoiseDec}
\left.\partial_t\hat{\rho}_r\right|_{\mathrm{dec}}= -\frac{1}{\hbar}\int_{0}^{t-t_i} d\tau \mathcal{N}(t;t-\tau)\left[\hat{x},\left[\hat{x},\hat{\rho}_{r}(t)\right]\right]\,.
\end{align}
The decoherence kernel in Eq.~\eqref{AEQ:DecKernel} for the problem at hand can also be computed more directly, without referring to the influence functional formalism, as it will be shown later. However, computing decoherence effects via Eq.~\eqref{AEQ:NoiseDec} has the advantage of discussing any possible connection between the Casimir force and decoherence in a straightforward way. Therefore, in what follows, the decoherence kernel $\mathcal{D}$ is derived using both approaches. 

\subsection{Decoherence}  
In the presence of parallel conducting plates, the freely evolved operator $\hat{\mathbf{\Pi}}$, same as the transverse electric field operator of the free EM field, is given by \cite{Barton1991, Hsiang2006} 
\begin{align}\label{AEQ:ElecField}
&\hat{\mathbf{\Pi}}(\x_{\parallel},x,t) =\nonumber\\
& -i\sqrt{\frac{2\hbar}{\epsilon_0L}}\sum_{n=0}^{\infty}f(n)\int \frac{d^2 k_{\parallel}}{2\pi}\sqrt{\frac{\omega_{n}}{2}} e^{i(\k_{\parallel}\cdot \x_{\parallel}-\omega_{n}t)}\times\nonumber\\
&\left[\hat{a}_{1}(\k_{\parallel},n)(\hat{\k}_{\parallel}\times\hat{\x})\sin(\frac{n\pi (x+L/2)}{L})+ \right.\nonumber\\
&\left.\ \hat{a}_{2}(\k_{\parallel},n)\left\lbrace \frac{i\hat{\k}_{\parallel}n\pi c}{\omega_{n}L}\sin(\frac{n\pi (x+L/2)}{L})\right.\right.\nonumber\\
&\left.\left.\qquad\qquad\quad-\frac{\hat{\x} k_{\parallel}c}{\omega_{n}}\cos(\frac{n\pi (x+L/2)}{L})\right\rbrace\right]+\mathrm{HC}\,.
\end{align}
Here, $f(n)$ is a weighing function such that $f(0) = 1/\sqrt{2}$ and $f(n)=1\,\forall n>0$ \cite{Casimir1948}, $\hat{\k}_{\parallel}$ is a unit vector along the plates, ${\omega^2_{n}/c^2 := k^2_{\parallel}+n^2\pi^2/L^2}$, and $\hat{a}_1$, $\hat{a}_2$ are the annihilation operators corresponding to the two modes of the EM field. It can also be seen that $\hat{\mathbf{\Pi}}_{\parallel}=0$ along the plates at $x=\pm L/2$, thus satisfying the appropriate boundary conditions. 

We are mainly interested in studying decoherence for the electron in a superposition along the $\hat{\x}$ axis. Thus, only the $\hat{\x}$ component of $\hat{\mathbf{\Pi}}$ enters the noise kernel in Eq.~\eqref{AEQ:NoiseDec}. For electron superpositions  prepared near the origin, on length scales $|x-x'|\ll L$ (superpositions over a distance $L$ will also be discussed later), the two-point correlation $\bra{0}\hat{\Pi}^x(1)\hat{\Pi}^x(2)\ket{0}$ is given by
\begin{align}
&\bra{0}\hat{\Pi}^x(1)\hat{\Pi}^x(2)\ket{0} =\nonumber\\ &\frac{\hbar c^2}{\epsilon_0 L}\sum_{n=-\infty}^{\infty}\int \frac{d^2k_{\parallel}}{(2\pi)^2}\frac{k^2_{y}+k^2_{z}}{2\omega_{n}}\cos^2(n\pi/2)e^{i(\k_{\parallel}\cdot \Delta\x_{\parallel}-\omega_{n}\tau)}
\end{align}
where $\Delta{\x}_{\parallel}:={\x}^{(1)}_{\parallel}-{\x}^{(2)}_{\parallel}$ and  $\tau:= t^{(1)}-t^{(2)}$. The surface integral above can be turned into a volume integral by introducing $\delta_n(k_x):=\delta(k_x-n\pi/L)$ such that
\begin{align}
&\bra{0}\hat{\Pi}^x(1)\hat{\Pi}^x(2)\ket{0} =\nonumber\\ &\frac{\hbar c^2}{\epsilon_0 L}\sum_{n\text{  even}}\int \frac{d^3k}{(2\pi)^2}\frac{k^2-k^2_{x}}{2\omega}e^{i(\k_{\parallel}\cdot \Delta\x_{\parallel}-\omega\tau)}\delta_n(k_x)\,.
\end{align}
Using the properties of the Dirac comb,
\begin{align}\label{AEQ:DiracComb}
\sum_{n\text{ even}} \delta_{n} = \sum_{n=-\infty}^{\infty}\delta_{2n} = \frac{L}{2\pi}\sum_{m=-\infty}^{\infty}e^{imk_x L}\,,
\end{align}
the two-point correlation becomes
\begin{align}\label{AEQ:TwoPoint_Dipole}
&\bra{0}\hat{\Pi}^x(1)\hat{\Pi}^x(2)\ket{0} =\nonumber\\
&\frac{\hbar c^2}{2\pi\epsilon_0}\sum_{m=-\infty}^{\infty}\hat{\Box}_m\int \frac{d^3k }{(2\pi)^22\omega}e^{i(\k\cdot \Delta\x_{m}-\omega\tau)} e^{-k/k_{\mathrm{max}}}\,,
\end{align}
where ${\Delta \x_{m}}:= \lbrace x_{m},\Delta \x_{\parallel}\rbrace$, ${x_{m}:=m L}$, $\hat{\Box}_{m}:= -\partial^2_{\tau}/c^2+\partial^2_{x_{m}}$, and the UV cutoff has been introduced in the calculations by inserting $\exp(-k/k_{\mathrm{max}})$ inside the integral. Further, the dependence on $\Delta \x_{\parallel}$ can be ignored not only because we are mainly interested in the dynamics along the $\x$ axis, where $\Delta \x_{\parallel}=0$, but also because $\Delta \x_{\parallel}\ll c\tau$ for a nonrelativistic particle. The integral can now be easily evaluated and gives
\begin{align}
\bra{0}\hat{\Pi}^x(1)\hat{\Pi}^x(2)\ket{0}=\frac{\hbar c}{\pi^2\epsilon_0}\sum_{m}\frac{1}{(m^2 L^2-c^2 (\tau-i\epsilon)^2)^2}\,,
\end{align}
where $\epsilon := 1/(k_{\mathrm{max}}c)$. The noise kernel is thus given by
\begin{align}\label{AEQ:NoiseKernel}
\mathcal{N}(\tau) = \frac{e^2 c}{2\pi^2\epsilon_0 }\sum_{m=-\infty}^{\infty}\frac{1}{(m^2 L^2-c^2 (\tau-i\epsilon)^2)^2}+ \text{HC}\,.
\end{align}
Some important observations can already be made at this stage. The $m=0$ term in the summation is independent of $L$ and is the only piece that survives in the limit ${L\to\infty}$. It must therefore correspond to the noise kernel $\mathcal{N}_0$ of empty space without any conducting plates.  Indeed, it can be seen that the $m=0$ term matches the noise kernel derived in \cite{Gundhi:2023vjs} (cf.~Eqs. (47) - (49) therein). When the acceleration of the charged particle due to an external potential can be neglected, the suppression of the off-diagonal elements of the density matrix due to $\mathcal{N}_0$ is false \cite{Diosi1995, Gundhi:2023vjs}. It is thus not relevant to observable decoherence effects. Therefore, the observationally relevant part $\mathcal{N}_{\mathrm{ob}}(\tau)$ is given by
\begin{align}\label{AEQ:Nobs}
\mathcal{N}_{\mathrm{ob}}(\tau) = \frac{e^2 c}{\epsilon_0 \pi^2}\sum_{m=1}^{\infty}\frac{1}{(m^2 L^2-c^2 (\tau-i\epsilon)^2)^2}+ \text{HC}\,.
\end{align}
Another interesting aspect emerges if we consider the large $L$ limit or, more correctly, the limit $c\tau\ll L$ which is given by
\begin{align}\label{AEQ:NCasimir}
\mathcal{N}_{\mathrm{Cas}} \overset{}{=}\frac{16 e^2}{\hbar \epsilon_0}\times\frac{\pi^2\hbar c}{720 L^4}\,.
\end{align}
The expression is written in a suggestive way since $\pi^2 \hbar c/(720 L^4)$ is nothing but the Casimir energy density (cf.~pg. 13 in \cite{Milton2001}). One might therefore attribute the leading order contribution to decoherence to the Casimir effect. This, however, should not be done. If expression~\eqref{AEQ:NCasimir} is used in Eq.~\eqref{AEQ:NoiseDec}, the off-diagonal elements of the density matrix would be suppressed as
\begin{align}\label{AEQ:DecCas}
\bra{x'}\hat{\rho}_r(t)\ket{x} \overset{\text{Cas.}}{=} \bra{x'}\hat{\rho}_r(0)\ket{x}\exp{\frac{-4\pi^3\alpha c^2t^2(x-x')^2}{90L^4}}\,.
\end{align}
We see that within the validity of the $1/L$ expansion of $\mathcal{N}_{\mathrm{ob}}$, i.e. $ct\ll L$, there will be practically no loss of coherence over any length scale $|x-x'|\leq L$. One might, however, extrapolate the consequences  to the time domain $ct\gg L$ and conclude that eventually coherence will be lost over time due to the plates. This should not be done, since the behavior of the full noise kernel is very different from the leading order term. 

To see this, the full expression for $\mathcal{N}_{\mathrm{ob}}$ in Eq.~\eqref{AEQ:Nobs} needs to be integrated. Doing that, the evolution of the off-diagonal elements, and thus the decoherence kernel (Eq.~\eqref{AEQ:DecKernel}), is obtained to be
\begin{align}\label{eq:Decoherence}
\rho_r(x',x,t)&=\exp\left(-\frac{(x'-x)^2}{\hbar}\mathcal{N}_2(t)\right)\rho_r(x',x,0)\,,\nonumber\\
\mathcal{N}_2(t)&:=\int_{0}^{t}dt'\int_{0}^{t'}d\tau\mathcal{N}_{\mathrm{ob}}(\tau)\nonumber\\
& =\sum_{m=1}^{\infty}\frac{e^2}{2\pi^2\epsilon_0m^3L^3}\times t\ln(\left|\frac{mL+ct}{mL-ct}\right|)\,,
\end{align}
where the limit $\epsilon\to 0$ has been taken at the end. Before going into the properties of the decoherence kernel $\mathcal{D}$, it is instructive to derive it using a more direct approach. As mentioned in Sec.~\ref{Sec:ImageCharges}, for a particle held in a superposition of $x$ and $x'$, $\mathcal{D}(x,x',t)$ is given by the overlap of the environmental states corresponding to the two particle locations. Given the interaction Hamiltonian in Eq.~\eqref{AEQ:Hint}, to leading order in $e$, the state of the environment (in the interaction picture) is given by
\begin{align}\label{AEQ:CoherentState}
\ket{\mathcal{E}(x)}_{t} = \exp(\frac{-i e x}{\hbar}\int_{0}^{t} dt'\hat{\Pi}^{x}(x,t') )\ket{0}\,.
\end{align}
Since the operator $\hat{{\Pi}}^{x}$ given in Eq.~\eqref{AEQ:ElecField} is simply a linear sum of creation and annihilation operators,  $\ket{\mathcal{E}(x)}_{t}$ is a coherent state $\ket{\alpha(x,t)}$. To work out the overlap between the coherent states corresponding to different particle locations, the use of dipole approximation is well justified  if the superpositions are prepared around $x=0$. Therefore 
\begin{align}
&\bra{\alpha(x',t)}\ket{\alpha(x,t)} = \prod_{\k_{\parallel}}\prod_{n}\bra{\alpha_{\k_{\parallel} n}(x',t)}\ket{\alpha_{\k_{\parallel} n}(x,t)}=\nonumber\\
&\exp\left\lbrace\frac{-e^2(x'-x)^2}{4\pi^2\hbar \epsilon_0L}\sum_{n=-\infty}^{\infty}\int d\k_{\parallel}\times\right.\nonumber\\
&\left.\qquad\times\frac{(k_{\parallel }c)^2\cos^2(n\pi/2)\sin^2(\omega_{n}t/2)}{ \omega^3_{n}}\right\rbrace\,,
\end{align}
where, the relation for the coherent states,
\begin{align}
\bra{\beta}\ket{\alpha} = \exp(-(|\beta|^2+|\alpha|^2-2\beta^*\alpha)/2)\,,
\end{align}
has been used. Using the Dirac comb~\eqref{AEQ:DiracComb}, the integral above can be evaluated exactly and gives
\begin{align}\label{AEQ:DecKernelPlates}
&\mathcal{D}(x,x',t) = \nonumber\\
&\exp{\frac{-e^2(x'-x)^2}{2\pi^2\hbar\epsilon_0 L^3}\sum_{m=1}^{\infty}\frac{t}{m^3}\ln(\left|\frac{mL+ct}{mL-ct}\right|)}\,,
\end{align}
where again, the contribution from the $m=0$ term has been discarded for the reasons described before. We see that the expression for the decoherence kernel is indeed the same when derived using the influence functional formalism (Eq.~\eqref{eq:Decoherence}), or when it is obtained by computing the overlap between the environmental states correlated to different electron positions.
\begin{figure}[!ht]
\centering
\includegraphics[width=0.70\linewidth]{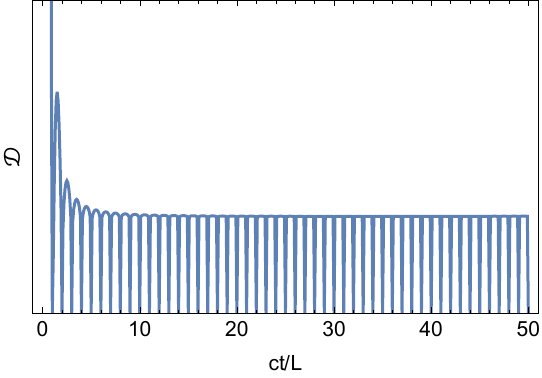}
\caption{Time evolution of the decoherence kernel showing suppression of the off-diagonal elements at times $t=mL/c$, and an otherwise constant curve for $ct\gg L$.}\label{Pic:DecKernel}
\end{figure}

\section{Initial jolt}\label{Sec:InitialJolt}
In this section it will be argued that all the features in $\mathcal{D}$, as depicted in Fig.~\ref{Pic:DecKernel}, are due to the sudden switching on of the S-E interaction. These effects should not be considered physical,  since in a realistic scenario there is no such sharp instant of time at which the electron starts interacting with the EM field vacuum. 

A sudden switching on of the S-E interaction results in a disturbance in the state of the environment, which is sometimes referred to as a \textit{jolt}  (cf.~\cite{HuPazZang1992}, the references therein, and \cite{Agon2017}). In general, after the initial jolt has passed away, an equilibrium between the system and the environment is established. However, in the setup that is being considered in this work, part of the initial jolt would always influence the dynamics and the S-E equilibrium is never established. This can be seen from the periodic fall in $\mathcal{D}$ at times $t_m= mL/c$.  

A possible explanation of these features is that the jolt in the EM field at time $t=0$ propagates, gets reflected by the plates, and arrives at the location of the electron again. Since the superpositions are assumed to be near the center, this jolt would arrive at the electron location again at times $t_m$, resulting in a temporary fall in coherence.\footnote{Sudden switching on of the S-E interaction might also lead to emission of real photons. However, possible irreversible decoherence due to scattering of such photons is a fourth order effect in $e$  and would thus be absent in $\mathcal{D}(x,x',t)$ shown here.} 

The same features can also be equivalently described in terms of the image charges.  It can be seen from Eq.~\eqref{AEQ:ImageLocation} that for an electron located at the center $x=0$, the nearest images are at a distance $L$, the next nearest at $2L$, and so on. As the initial jolt is sourced by the electron at the center, the images must source a counterjolt in order to maintain the boundary conditions. That is, by the time the jolt from the electron arrives at one of the plates, a counterjolt must arrive at the same plate from the nearest image to nullify the electric field along the plate. Subsequently, the jolt from the nearest images reaches the electron at $t=L/c$,  from the next nearest images  at $t=2L/c$ and so on,\footnote{The method of images can be applied not only to electrostatics, but also in dealing with the radiation field. For instance, the well known Casimir force was derived by explicitly referring to images in  \cite{Brown1969}.} explaining the fall in $\mathcal{D}$ at times $t_m$.

The reasoning above indicates that the sharp falls in the off-diagonal elements would not be present in the absence of a jolt. Therefore, in such a more realistic scenario, the decoherence kernel $\mathcal{D}$ would then only have a steady value, given by the constant upper bound in Fig.~\ref{Pic:DecKernel}. This asymptotic value can be calculated analytically. If $ct\gg L$, the main contribution to $\mathcal{D}$ comes from the modes for which $mL\ll ct$, since the contribution of the higher modes is greatly suppressed due to the $1/mL$ dependence. In this limit $\ln\lbrace(1+mL/ct)/(1-mL/ct)\rbrace\approx mL/ct$ and therefore
\begin{align}\label{AEQ:AsymptoticCurve}
\mathcal{D}(x,x',t)\approx \exp{\frac{-2\alpha (x'-x)^2}{\pi L^2}\sum_{m=1}^{\infty}\frac{1}{m^2}}\,.
\end{align}
The asymptotic expression for the coherence length is rather simple and involves both the fine structure constant $\alpha$ and the separation length $L$. Further, since $\sum_{m} 1/m^2 = \pi^2/6$, the off-diagonal elements would be unaffected over length scales $L/\sqrt{\alpha}$,\footnote{Although on such length scales the validity of the dipole approximation can be questioned. However, for any superpositions on scales much less than $L$, it is clear that no coherence is lost over time. The case of wider superpositions will also be discussed explicitly later.} which basically means no decoherence at all. 

Therefore, it is important to realize that the environment under consideration does not have a typical behavior. For instance, in the  environment of thermal photons, the off-diagonal elements decay exponentially in time (ignoring dissipation, cf.~Fig.~3.8 in \cite{KieferDecoherence}). Thus, the extrapolation of Eq.~\eqref{AEQ:DecCas} in thinking that the Casimir force leads to a continuous irreversible loss of coherence, like ordinary environments, does not hold. 

It was shown in \cite{Gundhi:2023vjs} that if one considers a time dependent coupling $q(t)$ with the environment, i.e., $q(t)=-ef(t)$ such that $f(t)=1$ for most of the dynamics between the initial time $t=0$ and the final time $t=T$, while $f(0)=f(T)=0$, the  noise kernel in Eq.~\eqref{AEQ:NoiseDec}, and defined in Eq.~\eqref{AEQ:Kernels}, transforms as $\mathcal{N}\rightarrow\tilde{\mathcal{N}}$, with  
\begin{align}
\tilde{\mathcal{N}} = f(t_1)f(t_2)\mathcal{N}(t_1;t_2) =  f(t_1)f(t_2)\mathcal{N}(t_1-t_2)\,.
\end{align}
It was further shown at a very general level that if $\mathcal{N}_2$ approaches a constant value on some timescale, which in this case is $L/c$, and if the S-E interaction is switched on adiabatically over this  timescale, then  $\mathcal{N}_2$ transforms as  $\mathcal{N}_2\to\tilde{\mathcal{N}}_2$  (cf.~Eq.~(75) in \cite{Gundhi:2023vjs}) with
\begin{align}\label{AEQ:falsedec}
\tilde{\mathcal{N}}_2(T)=\frac{\mathcal{N}_2}{2}\left(f^2(0)+f^2(T)\right) = 0\,,
\end{align}
implying no irreversible decoherence. These considerations imply that in a typical scenario in which the S-E interaction is not switched on suddenly, there would be no irreversible loss of coherence at the level of the electron and thus no loss in the fringe contrast in a double slit type of experiment. The reasoning and the arguments presented in this section are confirmed in Sec.~\ref{Sec:AdiabaticSwitching}, where the decoherence kernel is computed after adiabatically switching on the S-E interaction.

\section{Superpositions over larger length scales}
For particle superpositions near the center, the analysis shows that there is no loss in coherence over time and that the features in $\mathcal{D}(x,x',t)$ are solely due to the sudden switching on of the interactions. One may still question the validity of these conclusions for superpositions prepared over large length scales. It should first be clear that when considering superpositions along the $\x$ axis, the results will not be affected by the $y$ and $z$ dependence of the noise kernel. This is because the dependence on the difference between the $y$ and $z$ coordinates, of the nonrelativistic particle superpositions, can be ignored since $\Delta \x_{\parallel}\ll c\tau$. This can be seen easily from Eq.~\eqref{AEQ:TwoPoint_Dipole}.\footnote{See also the discussion in Sec. 2.1.5 (and the footnote below) of \cite{Tannoudji1982}, where it is mentioned that corrections to the dipole approximation are only relevant when relativistic effects become important.} 

With this in mind, the decoherence kernel for the electron in a superposition over large length scales, $x=-L/2$ and $x'=L/2$, is now computed. In this case, the two-point correlation is given by 
\begin{align}
&\bra{0}\hat{\Pi}^x(-L/2,t_1)\hat{\Pi}^x(L/2,t_2)\ket{0} =\nonumber\\ &\frac{\hbar c^2}{\epsilon_0 L}\sum_{n=-\infty}^{\infty}\int \frac{d^2k_{\parallel}}{(2\pi)^2}\frac{k^2_{y}+k^2_{z}}{2\omega_{n}}(-1)^ne^{i(\k_{\parallel}\cdot \Delta\x_{\parallel}-\omega_{n}\tau)}\,.
\end{align}
As before,  the surface integral can be converted into a volume integral by inserting a Dirac comb, and then summing over $\sum_{n}(-1)^n\delta_{n}$. This sum can be written as
\begin{align}\label{AEQ:DiracComb_LargeSep}
\sum_{n=-\infty}^{\infty} (-1)^n\delta_{n} = 2\sum_{n=-\infty}^{\infty} \delta_{2n} - \sum_{n=-\infty}^{\infty} \delta_{n}\,.
\end{align}
The identity involving the Dirac comb can now be applied to both terms on the rhs of the equation above. With little algebra, it is easy to see that it gives
\begin{align}
\mathcal{N}(\tau) = &\frac{e^2 c}{\epsilon_0\pi^2 }\sum_{m=-\infty}^{\infty}\frac{1}{(m^2 L^2-c^2 (\tau-i\epsilon)^2)^2}+ \text{HC}\nonumber\\
&\frac{-e^2 c}{\epsilon_0\pi^2 }\sum_{m=-\infty}^{\infty}\frac{1}{(m^2 (2L)^2-c^2 (\tau-i\epsilon)^2)^2}+\mathrm{HC}\,.
\end{align}
 The decoherence kernel is then immediately obtained to be
\begin{align}\label{AEQ:LargeDistance}
&\mathcal{D}(-L/2,L/2,t) = \nonumber\\
&\exp\left\lbrace\frac{-e^2(x'-x)^2}{\pi^2\hbar\epsilon_0}\sum_{m=1}^{\infty}\frac{t}{m^3}\times\right.\nonumber\\
&\left.\left[\frac{1}{L^3}\ln(\left|\frac{mL+ct}{mL-ct}\right|)-\frac{1}{(2L)^3}\ln(\left|\frac{2mL+ct}{2mL-ct}\right|) \right]\right\rbrace\,.
\end{align}
Consistent with the description above, it can be seen that there will be falls in coherence not only at time intervals $t=2mL/c$,  since it takes the time $2L/c$ for the jolt sourced near one of the plates to return to the same plate, but also at time intervals $t=mL/c$, since the jolt sourced from $x=-L/2$ can temporarily become correlated to the electron at $x'=L/2$.  Again, even for the largest possible superposition, one sees that all the features in the off-diagonal elements can be ascribed to the sudden switching on of the interaction. The same and consistent conclusions are reached if $\mathcal{D}(-L/2,0,t)$ or $\mathcal{D}(0,L/2,t)$ is computed.

\section{Adiabatic switching on}\label{Sec:AdiabaticSwitching}
To confirm the physical interpretation, one can compute the off-diagonal elements of the density matrix, but this time after adiabatically switching on the S-E interaction. This can be modeled by evolving the state of the environment as
\begin{align}\label{AEQ:CoherentStateAdiabatic}
\ket{\mathcal{E}(x)}_t^{\mathrm{ad}} = \exp\left\lbrace\frac{-i e x}{\hbar}\int_{0}^{t} dt'\hat{\Pi}^{x}(x,t') \left(1-e^{-t'/T}\right)\right\rbrace\ket{0}\,.
\end{align}
Here, $T$ represents the timescale over which the S-E interaction is switched on and $t$ is some late time at which the interaction is fully switched on. After computing the integral, one should first take the limit $t\gg T$ and then the limit $T\to\infty$. If the limits are taken in the opposite order, the S-E interaction would never be switched on. After the interaction is fully switched on at some time $t$, the decoherence kernel is given by
\begin{align}\label{AEQ:OverlapAdiabatic}
&\mathcal{D}^{\mathrm{ad}}(x,x',t)=\bra{\mathcal{E}(x')}\ket{\mathcal{E}(x)}^{\mathrm{ad}}_t =\nonumber\\ 
&\exp\left\lbrace\frac{-e^2(x'-x)^2}{16\pi^2\hbar \epsilon_0L}\sum_{n=-\infty}^{\infty}\int d\k_{\parallel}\frac{(k_{\parallel }c)^2\cos^2(n\pi/2)}{ \omega^3_{n}}\right\rbrace\,.
\end{align}
Here, the calculations are performed for the case where the superpositions are prepared near the center. The main effect of a sudden jolt is not related to the length scale over which the superpositions are prepared and the conclusions of adiabatically switching on the interaction would also apply to $\mathcal{D}(L/2,-L/2,t)$.

It is already interesting to notice that $\mathcal{D}$ in Eq.~\eqref{AEQ:OverlapAdiabatic} does not depend on time $t$, indicating that the  oscillations in Fig.~\ref{Pic:DecKernel} do not appear in the absence of a sudden jolt. As before, the integral can be computed by inserting a Dirac comb and using the identity~\eqref{AEQ:DiracComb}. It gives
\begin{align}
&\mathcal{D}^{\mathrm{ad}}(x,x',t)=\nonumber\\ 
&\exp\left\lbrace\frac{-e^2(x'-x)^2}{8\pi^2\hbar c\epsilon_0}\times\right.\nonumber\\
&\left.\times\sum_{m=1}^{\infty}\int_{0}^{\infty} dk\,k\int_{0}^{\pi} d\theta \sin \theta (1-\cos^2\theta)e^{imLk\cos\theta}\right\rbrace\nonumber\\
&= \exp\left\lbrace\frac{-2\alpha (x'-x)^2}{\pi L^2}\sum_{m=1}^{\infty} \frac{1}{m^2}\right\rbrace\,.
\end{align}
It can be clearly seen that the oscillations have disappeared from the decoherence kernel altogether (compared to Eq.~\eqref{AEQ:DecKernelPlates}) and that its value is the same as that of the asymptotic curve in Fig.~\ref{Pic:DecKernel}. The off-diagonal elements are not suppressed over any length scale $|x-x'|\leq L$. What is even more important to notice is that after the S-E interaction is fully  switched on, the overlap between the environmental states is stationary in time. Thus, the vacuum of the interacting radiation field effectively acts like an environment which is only correlated to the position of the electron. The situation then becomes identical to the one described in Sec.~\ref{Sec:ImageCharges}. This implies that there would be no loss in the fringe contrast due to vacuum fluctuations,  as they are not able to resolve the two paths which end at the same point on the detector screen.

\section{Comparison with previous works}
In this section a comparison is made with previous works  \cite{Ford1993,Mazzitelli2003,Hsiang2006} where it was shown that vacuum fluctuations in the presence of a conducting plate can lead to finite decoherence, contrary to the conclusion reached in this work.

Ref.~\cite{Ford1993} studies decoherence due to vacuum fluctuations in the presence of a single conducting plate. However, since Eq.~(57) in Ref.~\cite{Ford1993} resembles  Eqs.~\eqref{AEQ:DecKernelPlates} and~\eqref{AEQ:LargeDistance} in the present work (though it is not the same), the analysis in \cite{Ford1993} is most likely performed without switching on the S-E interaction adiabatically, and therefore the results obtained therein represent decoherence due to the initial jolt described in Sec.~\ref{Sec:InitialJolt}. 

The standard assumption of starting with $\hat{\rho}(t_i) = \hat{\rho}_{s}(t_i)\otimes\ket{0}\bra{0}$ cannot be made when studying interaction with vacuum fluctuations, since there is no way to \textit{avoid} the vacuum. One might work with this assumption, however, after switching off the S-E interaction by hand at the initial time and then adiabatically switching it on. As explained in the last section, when this is done, there would be no decoherence due to vacuum fluctuations in the presence of conducting plates. Having two parallel conducting plates introduces infinite number of image charges instead of one (as in the case of a single conducting plate). Thus, it is reasonable to assume that the conclusions reached in this article would also hold for the case in which only a single conducting plate is present \cite{Ford1993}.

Refs.~\cite{Mazzitelli2003,Hsiang2006} also obtain a finite value of decoherence in the presence of a conducting plate, although their result is  different to the one in \cite{Ford1993}. However, the explanation for the nonzero value found in \cite{Mazzitelli2003,Hsiang2006} is different and a bit more subtle. 

In order to avoid misinterpretations concerning the decay of the off-diagonal elements $\rho_r(x',x,t)$, such as the one described in Sec.~\ref{Sec:ImageCharges}, one might analyze if the environment can decohere a superposition of different trajectories rather than different positions. However, even such an analysis, in general, is not free of misinterpretations.

According to the axes-convention used in Fig.~\ref{Images}, \cite{Mazzitelli2003,Hsiang2006} study decoherence for a superposition of trajectories $(t,R e^{-t^2/T^2},v_yt,0)$ and  $(t,-R e^{-t^2/T^2},v_yt,0)$. See, for example, the discussion below Eq.~(11) in \cite{Mazzitelli2003} and above Eq.~(27) in \cite{Hsiang2006}. Since the acceleration of the electron along the two paths is finite, of the order $R/T^2$, in such an analysis, in addition to any possible decoherence due to vacuum fluctuations, one also gets decoherence due to photons emitted by the accelerating electron. In fact, the finite decoherence obtained in the two works is due to bremsstrahlung induced by the chosen  background trajectories whose superposition is analyzed and \textit{not} due to vacuum fluctuations. The reasoning is based on the following estimate:

Decoherence due to bremsstrahlung, for an electron accelerated by an external harmonic potential, is given by 
\begin{align}
\dot{\hat{\rho}}_{r}(t)|_{\text{dec}} = -\frac{\alpha \Omega^3}{3 c^2}[\hat{x},[\hat{x},\hat{\rho}_r]]\,.
\end{align}
Here, only the double commutator term on the rhs of Eq.~\eqref{AEQ:Brem} is retained. This equation can be readily integrated to see that the off-diagonal elements decay as
\begin{align}\label{eq:BremDec}
\rho_{r}(x',x,t) = \exp{-\frac{\alpha\Omega^3 t (x'-x)^2}{3c^2}}\rho_{r}(x',x,0)\,.
\end{align}

Now, the main objective in \cite{Mazzitelli2003,Hsiang2006} is to study decoherence between two trajectories that diverge up to a distance $R$, and then converge back again on a timescale $T$. This can be equivalently achieved with the trajectories $(t,\pm R\sin{\Omega t},v_yt,0)$ where $t$ goes from $0\to \pi/\Omega$. In this way, decoherence due to bremsstrahlung over trajectories considered in \cite{Mazzitelli2003,Hsiang2006}  can be equivalently estimated using Eq.~\eqref{eq:BremDec}. Setting $\Omega  = 1/T$ and $(x'-x)^2 = R^2$, one obtains
\begin{align}\label{eq:DecBrem}
\rho_{r}(x',x,T) \sim \exp{-\frac{\alpha \pi R^2}{3c^2 T^2}}\rho_{r}(x',x,0)\,.
\end{align}

This estimate closely resembles the expression for decoherence $\exp{W_0}$,  $W_0\propto -e^2 R^2/(c^2T^2)$, found in \cite{Mazzitelli2003,Hsiang2006} (cf.~Eqs.~(23, 31) in \cite{Hsiang2006}). However, in \cite{Mazzitelli2003,Hsiang2006} this result is misinterpreted as decoherence due to vacuum fluctuations in empty space. To avoid decoherence due to photon emission, one could consider the adiabatic trajectories $T\to\infty$, so that the acceleration along the paths can be ignored. As expected, there is no decoherence in this limit.

According to \cite{Mazzitelli2003,Hsiang2006}, there is decoherence due to vacuum fluctuations even in the absence of conducting plates, which contradicts previous works \cite{Diosi1995, Unruh_Coherence, Gundhi:2023vjs}. Refs.~\cite{Mazzitelli2003,Hsiang2006} further show that the presence of plates may increase or reduce the magnitude of this decoherence. However, since \cite{Mazzitelli2003,Hsiang2006} use the same trajectories in their analyses of a conducting plate, what they are possibly computing are corrections to decoherence due to bremsstrahlung, induced by the acceleration of the electron  that is enforced by the chosen background paths. The same issue prevails for the case of two parallel conducting plates examined in \cite{Hsiang2006}.

\section{Conclusions} In this work, decoherence due to vacuum fluctuations of the EM field confined between parallel conducting plates is analyzed. It is argued that the loss in coherence  {which has been found in previous works is either due to the sudden switching on of the S-E interaction \cite{Ford1993}, or due to bremsstrahlung induced by the chosen background paths for which decoherence is analyzed \cite{Mazzitelli2003,Hsiang2006}. It is further shown in this work that decoherence in the presence of conducting plates is not a consequence of the nature of the vacuum fluctuations themselves, and thus not related to the Casimir force.

To study the time evolution of the reduced density matrix, it is standard practice, for technical convenience, to start from an initially uncorrelated S-E state. 
However, this initial condition, in general, is only justified when the interaction between the system and the environment is also switched off initially, since S-E interaction leads to S-E  correlations. 

Therefore, if one takes the vacuum state of the free EM field as the initial state of the environment, uncorrelated to the initial state of the electron, one should also switch off their interaction initially. It is well known from standard quantum mechanics that to go from the vacuum state of the free EM field (i.e. the bare vacuum), to the vacuum state of the interacting radiation field (which is the main subject of the analysis), the interaction must be switched on adiabatically. 
In other words, the gap between the starting point (dictated by technical convenience), and the actual physical situation of interest, must be bridged by adiabatically switching on the interaction. When this is done, it is shown in this work that the zero-point modes of the vacuum do not lead to any decoherence effects at the level of the electron. Decoherence due to zero-point modes might still be relevant in a situation where the charged particle suddenly enters and leaves a region confined between conducting plates. However, in typical scenarios, the only irreversible loss of coherence in the presence of parallel plates would be due to the effective Coulomb potential of all the infinite image charges, which accelerates the electron resulting in bremsstrahlung.

\section{Acknowledgements}
I thank Angelo Bassi and fellow group members for several discussions. This work was financially supported by the University of Trieste, INFN and EIC Pathfinder project QuCoM (GA No. 101046973). 

\bibliography{VacuumDecoherenceBib}{}
\bibliographystyle{apsrev4-2}
\end{document}